\documentclass[11pt]{article}
\usepackage{moriond,epsfig}

\def\be{\begin{equation}}
\def\ee{\end{equation}}
\def\bea{\begin{eqnarray}}
\def\eea{\end{eqnarray}}

\begin{document}
\vspace*{4cm}
\title{Heavy Quark Dynamics in Heavy Ion Reactions}

\author{ J.L. Nagle }

\address{Department of Physics, University of Colorado at Boulder,\\
Boulder, CO 80309, USA}

\maketitle\abstracts{
Collisions between heavy nuclei at the Relativistic Heavy Ion Collider 
liberate from the nuclear wavefunction of order 10,000 gluons, quarks and antiquarks.  
The system is dominated by gluons and up and down (anti) quarks.  Heavy quarks, though
having little effect on the overall equation of state, are critical as 
probes of the surrounding medium.  
We compare predictions from a scenario where the charm quarks escape 
the medium unaffected and fragment into hadrons in vacuum, and one
where the charm quarks are swallowed up by the medium and reflect the
overall system hydrodynamics.
}

\section{Introduction}

A critical question in need of resolution is how fast moving partons interact with
a medium of color charges.
We can study this question when hard scattered partons are embedded in a nucleus as in
deep inelastic scattering on nuclei ($e-A$) or in proton-nucleus reactions ($p-A$), or in
a energy dense system created in nucleus-nucleus reactions ($A-A$).
It has also been proposed that prior to hadronization,
partons traversing a color dense medium will radiate additional gluons.

At the Relativistic Heavy Ion Collider the conventional picture has been that the 
scattered quarks propagate transversely through
the medium and then fragment into hadrons outside in vacuum.  In the presence of
a dense gluonic medium, the parton scatters off of many color charges and emits significant
additional gluons.  In this scenario, the jet fragmentation is significantly
softer since the parton loses energy before hadronizing, thus suppressing high z hadron
formation.  This picture assumes no interaction of a formed color dipole or the later fully 
formed hadron with the medium.
At RHIC there have now been numerous observations of suppression of high $p_{T}$ hadrons
relative to factorization expectations scaled by the nuclear thickness function~\cite{phenix_highpt}.  


\section{Flavor Dependence}


Charm quarks are an ideal probe since their production is dominantly via gluon fusion 
and has a characteristic $Q^{2}$ bound from below by the mass of the $c\overline{c}$ pair.  
QCD is flavor independent and thus to first order one might expect the energy loss of charm
quarks to be the same as light up and down quarks in medium.

\begin{figure}
\psfig{figure=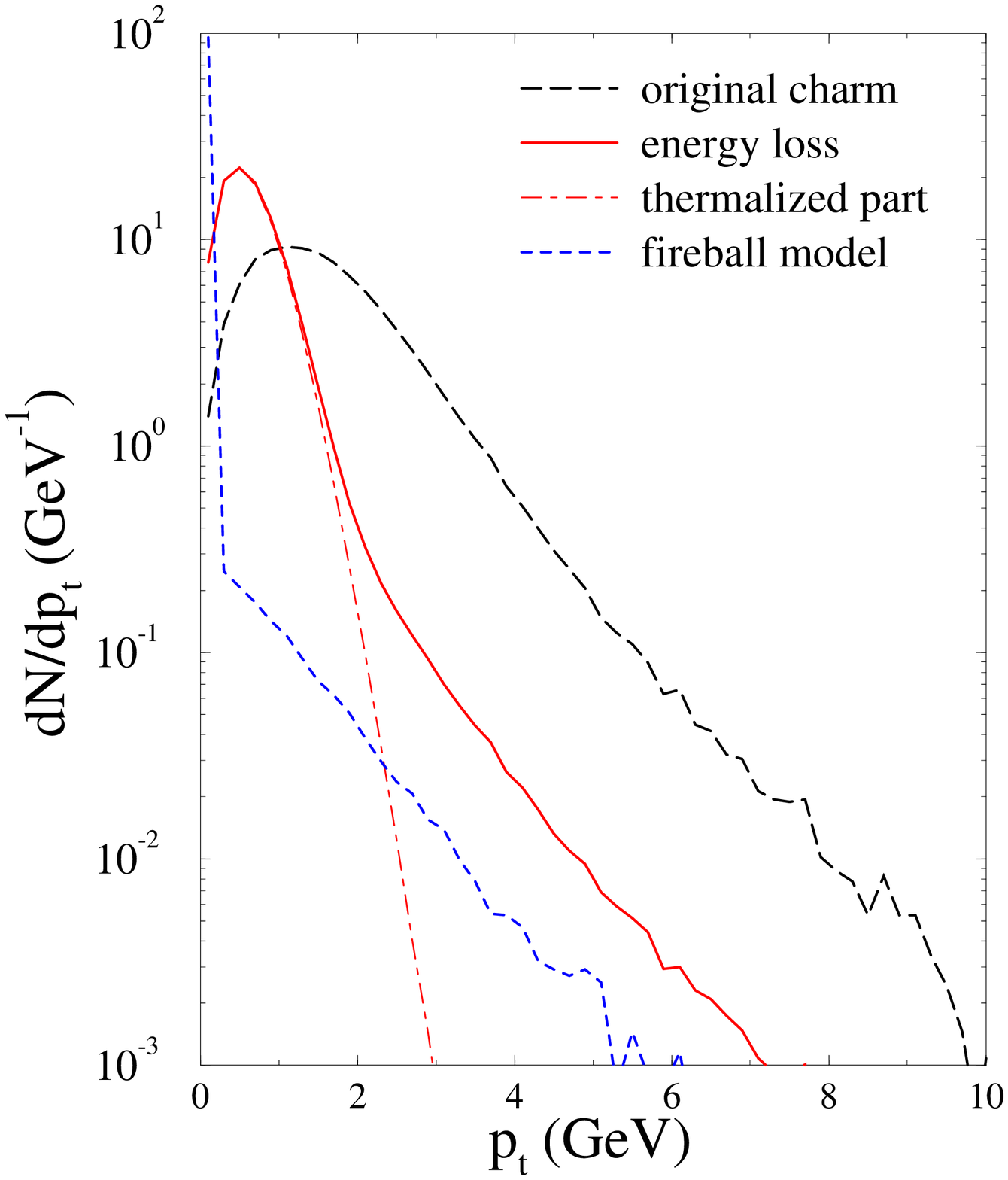,height=2.0in}
\psfig{figure=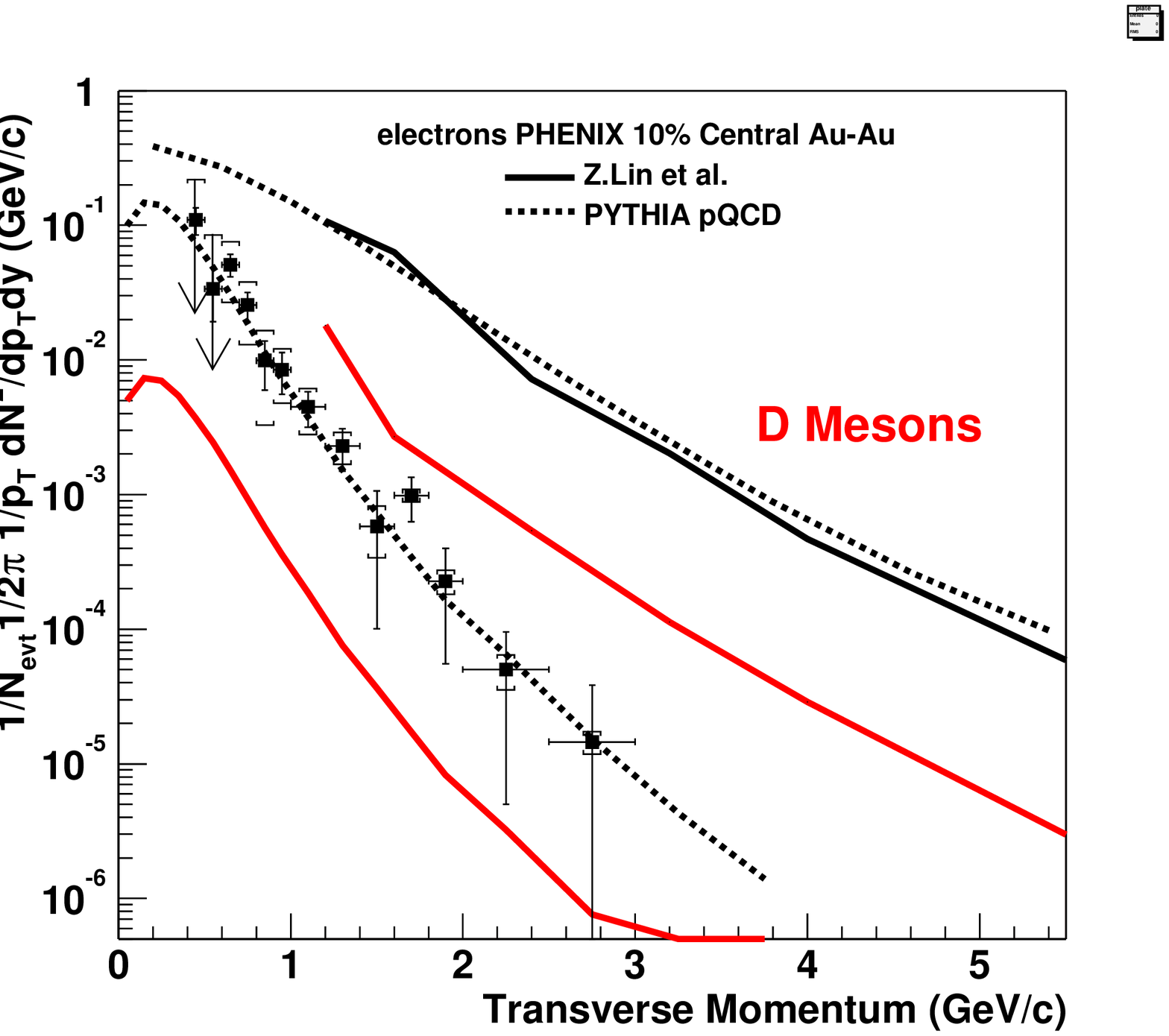,height=2.0in}
\caption{(Left Panel) Calculation of original charm production with and without $dE/dx = -2.0 GeV/fm$.  (Right Panel) PHENIX single electron data with calculation comparisons.
\label{fig:radish}}
\end{figure}

A calculation~\cite{lin1} includes a charm quark energy loss in medium ranging from 
$dE/dx = -0.5, -1.0, -2.0$ 
GeV/fm in Au-Au reactions at RHIC.  The $dE/dx = -2.0$ GeV/fm leads to
an order of magnitude suppression in intermediate $p_{T}$ charm quarks as shown in Figure 1.  
It is important to note that since
charm is a conserved quantum number and $c\overline{c}$ annihilation is infrequent,
charm quarks that lose energy must pile up at low $p_{T}$.
Charm quarks that lose all their energy are assigned a thermal distribution of momenta.
We have adjusted their calculation to employ a Peterson Fragmentation
function (rather than their $\delta(1-z)$ fragmentation function), input a 
charm cross section of 330 $\mu b$ for $\sqrt{s}$ = 130 GeV, and rescaled the result to match
the expected number of binary collisions for 10\% central Au-Au reactions.  
We show the D meson spectra and 
resulting single electron spectra from semi-leptonic decays in the right panel 
of Figure 1.  Note that we have not included the pile-up at the lowest $p_{T}$.  
We have also calculated the yield of D mesons and resulting 
single electrons in the PYTHIA model~\cite{pythia}.
The PYTHIA model has been tuned to match lower energy charm meson data and ISR data of ``prompt''
single electrons.  It predicts a proton-proton
$\sigma_{c\overline{c}} =$ 330 $\mu b$ and 650 $\mu b$ at 130 and 200 GeV, respectively~\cite{pythia}.
The uncertainty in these cross sections are of order $\pm$ 50\%.

The PHENIX experiment has reported first results in Au-Au collisions at 130 A GeV 
of ``prompt'' single electrons from charm and beauty meson decays~\cite{phenixe}.
The PHENIX data for central Au-Au reactions is also shown in the right panel of Figure 1.
The data disfavor large energy loss, though a smaller effect cannot be ruled
out due to the large systematic errors on the first year PHENIX data and the 
uncertainty in the charm cross section.
It is a crucial measurement to determine the total charm cross section in proton-proton
reactions at RHIC energies.

One proposed explanation for the lack of observed energy loss of heavy quarks, despite the large effect
implied by the light quarks and gluons, is the ``dead-cone'' effect.  Since a charm quark with 
$p_{T} \approx$ 3.0 Gev
has a velocity significantly less than the speed of light ($v \approx 0.92~c$), there will be a suppression
of forward radiated gluons~\cite{dima}.  
Recent work indicates that this effect is not sufficient
to reduce the induced gluon radiation to reconcile theory with data~\cite{mag}.  
However, it has been proposed that an additional
effect may reduce the zeroth order gluon radiation~\cite{mag}.  
This so called
Ter-Mikayelian effect essentially enhances the charm $p_{T}$ spectra via a hardening of the 
vacuum fragmentation
function by introducing a plasma cutoff frequency on the radiated gluons.  This effect is then roughly
canceled by the still present induced energy loss in the dense gluonic matter produced in the 
heavy ion collision.

\section{Charm Hydrodynamics}

We recently proposed an alternative explanation for the single electron data~\cite{batsouli}.
In the calculation with large energy loss~\cite{lin1}, charm quarks that were stopped in medium were simply
given a thermal momentum distribution with a temperature of 150 MeV.  However, in relativistic heavy ion
collisions we observe a strong positive feedback from the system.  
Large pressure built up in the earliest
partonic stage results in a large transverse velocity boost to the final state hadrons and
a large anisotropic flow in non-central reactions.  These features have been 
measured by all the RHIC experiments
for hadrons composed of up and down quarks (mesons and baryons), and also including strange hadrons (perhaps
even the omega baryon)~\cite{kolb}.

\begin{figure}
\psfig{figure=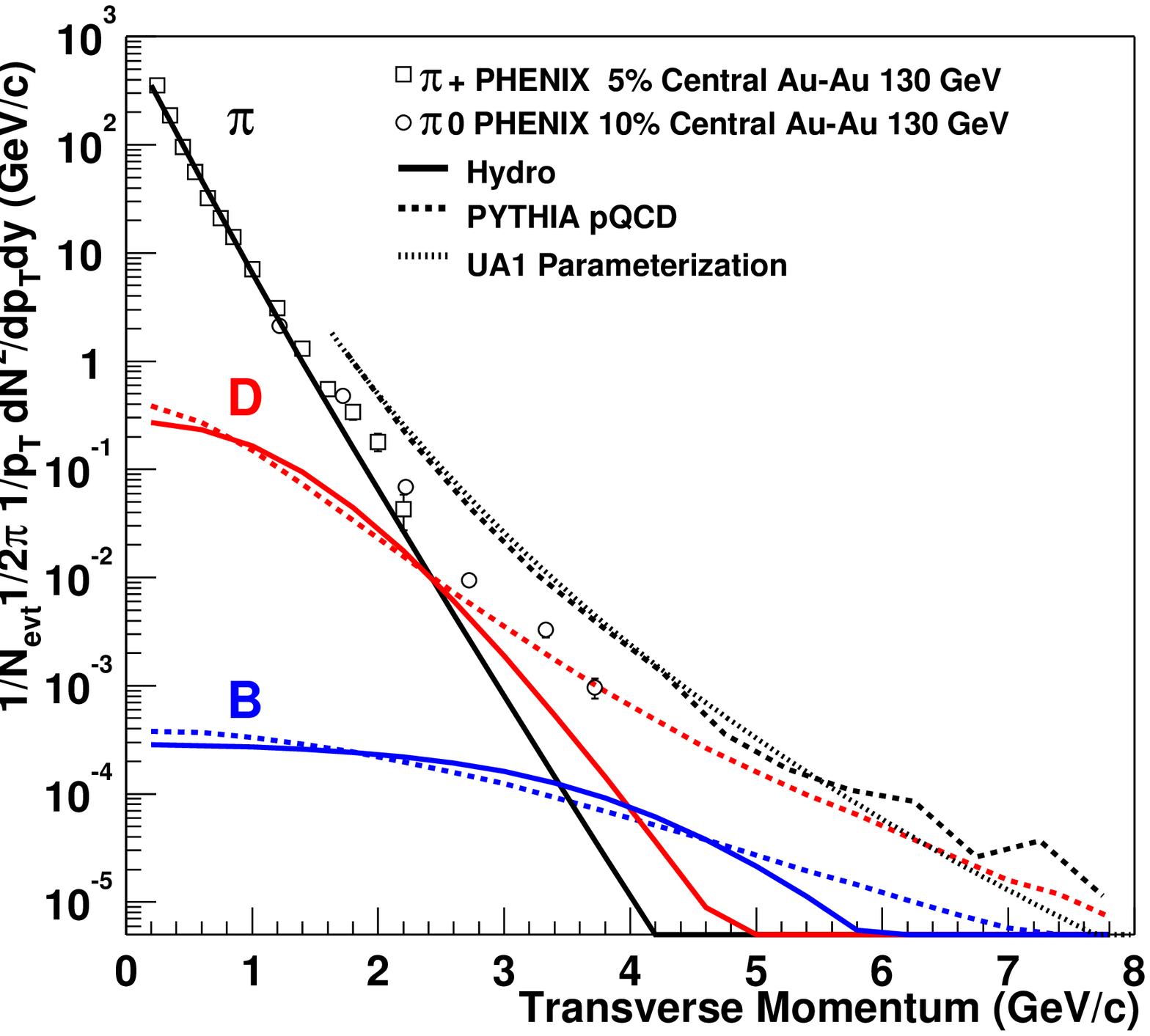,height=2.0in}
\psfig{figure=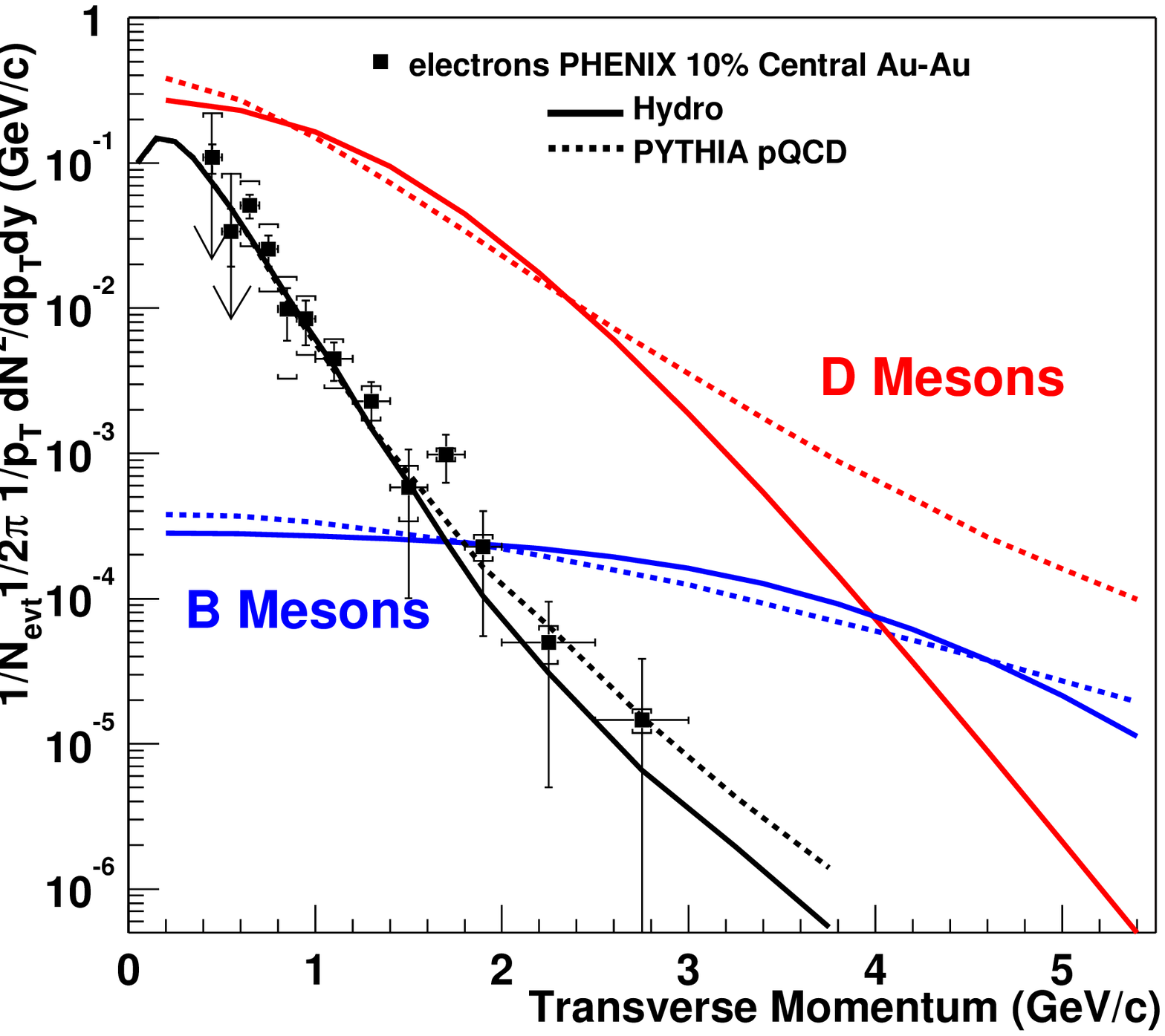,height=2.0in}
\caption{(Left Panel) $p_{T}$ spectra for central Au-Au reactions for 
$\pi, D, B$ mesons.  The solid
lines are from a hydrodynamic calculations and the dashed lines from pQCD 
inspired PYTHIA model.  (Right Panel) D 
and B mesons and electrons from hydrodynamics, 
PYTHIA and the PHENIX data.
\label{fig:radish}}
\end{figure}

We suggest that perhaps even if the charm quarks lose substantial energy in medium, they will then
re-scatter and receive the same transverse velocity boost as the other light quarks and gluons.  
Hydrodynamic models have been quite successful at describing a number of features in the final state
hadron spectra.  We employ a simple parameterization of the final momentum distribution in a 
hydrodynamic picture~\cite{flow}.
We fit the PHENIX measured $\pi, K, p$ transverse momentum spectra~\cite{phenix_hadrons} 
assuming a constant temperature of T=128 MeV and find hydrodynamic
parameters $\beta_{max}=0.65, 0.65, 0.63, 0.55, 0.25$ for Au-Au centralities of 0-5\%,
5-15\%, 15-30\%, 30-60\%, 60-92\%, respectively.  We can then calculate the transverse
momentum distribution for D and B mesons with no additional parameters.  Note that we normalize the
D and B integrals to agree with the PYTHIA total yield calculation.  Thus in the language of thermodynamics,
we have not assumed chemical equilibration of charm and beauty, but rather that whatever
heavy quarks are produced in initial hard scattering or gluon fusion reactions are conserved.

We show in the left panel of Figure 2 the transverse momentum distribution of $\pi$, D and B mesons within
the hydrodynamic approach (solid lines), and with the pQCD inspired PYTHIA model (dashed lines).
We also shown the PHENIX measured $\pi$ which agree with the hydrodynamic model at low $p_{T}$ and then 
diverge, though still showing the characteristic suppression relative to pQCD expectations.
The D and B mesons show that these two scenarios lead to very similar spectra out to 3 and 5 GeV respectively.
We show in the right panel of Figure 2 the resulting electron predictions from the hydrodynamic
model and PYTHIA.  There is an ambiguity between a scenario with a completely transparent system
with charm quarks fragmenting in vacuum (PYTHIA) and a scenario with a completely 
opaque system characterized by large energy loss and subsequent
re-scattering that boosts the heavy quarks to higher $p_{T}$.

Preliminary Au-Au data from PHENIX at $\sqrt{s}$ = 200 GeV~\cite{ralf} are also consistent with 
a transparent medium and no energy loss.  In particular, the charm production appears to scale with 
the number of binary collisions from the 10\% most central events ($N_{binary} = 975$) through
more peripheral 40-70\% events ($N_{binary}$ = 71).
Based on our hydrodynamic parameters, we predict
a relatively similar $p_{T}$ dependence of the electrons for centralities all the way out to 60\% central. 
Thus one would like to see the spectra shape for a more peripheral class of events where the
hydrodynamic velocity boost is much decreased.
It should be noted that for the most peripheral events, even the elliptic flow $v_{2}$ 
predicted by hydrodynamics significantly over-predicts the experimental data.  Therefore, there are
other indicators that hydrodynamics may not be applicable for very small volume systems.
In addition, in some high $p_{T}$ range we do not expect hydrodynamics to be valid, as has been 
experimentally observed for other hadrons.

We suggest that a measurement of the anisotropic flow ($v_{2}$) for charm mesons or $J/\psi$ would 
yield definitive evidence to discriminate these pictures.  If the medium is truly transparent
to charm quarks, the D meson distribution should have no correlation to the initial collision
reaction plane.  In a recent charm recombination picture, they set the range for charm hadron
$v_{2}$ by assuming all flow comes from the combined light quark and one in which the 
charm quark flows as well~\cite{molnar}.

\section{Summary}

It is critical to understand the hadronization process of hard scattered and created partons
in the heavy ion environment.  Comparisons between various nuclear systems provides crucial
information for developing a comprehensive picture of hadronization and in medium energy loss.
The charm and beauty quarks provide an additional handle on this process.  Current data
do not allow the discrimination of quite different dynamical scenarios leading to the 
final ``prompt'' electron distribution.  However, data in the near future should aid in
clarifying this picture.

\section*{Acknowledgments}

This proceedings is dedicated to the memory of my father John David Nagle.
We thank Sean Kelly for a careful reading of the proceedings.
JLN acknowledges support by the U.S. Department of Energy 
under Grant No. DE-FG02-00ER41152 and
the Alfred P. Sloan Foundation.

\end{document}